\documentclass[12pt]{iopart}
\usepackage{iopams}
\usepackage{longtable}
\usepackage{graphicx}
\usepackage{dcolumn}

\usepackage{epsfig,cite,colordvi}

\usepackage{url}
\usepackage[colorlinks=true,urlcolor=blue]{hyperref}
\usepackage{amssymb}
\usepackage{times}
\usepackage{makeidx}
\begin{document}

\title{Various Aspects of the Deformation Dependent Mass model of Nuclear Structure}

\author{D. Petrellis $^1$, D. Bonatsos $^2$, N. Minkov $^3$},

\address{$^1$ Department of Physics, University of Istanbul, 
34134 Vezneciler, Istanbul, Turkey}

\address{$^2$ Institute of Nuclear and Particle Physics, 
National Centre for Scientific Research "Demokritos," 
GR-15310 Aghia Paraskevi, Attiki, Greece}

\address{$^3$ Institute of Nuclear Research and Nuclear Energy, 
Bulgarian Academy of Sciences, 72 Tzarigrad Road, 1784 Sofia, Bulgaria}

\begin{abstract}
Recently, a variant of the Bohr Hamiltonian was proposed where the mass term is allowed to depend on the $\beta$ variable of nuclear deformation. Analytic solutions of this modified Hamiltonian have been obtained using the Davidson and the Kratzer potentials, by employing techniques from supersymmetric quantum mechanics. Apart from the new set of analytic solutions, the newly introduced Deformation-Dependent Mass (DDM) model offered a remedy to the problematic behaviour of the moment of inertia in the Bohr Hamiltonian, where it appears to increase proportionally to $\beta^2$. In the DDM model the moments of inertia increase at a much lower rate, in agreement with experimental data.	
The current work presents an application of the DDM-model suitable for the description of nuclei at the point of shape/phase transitions between vibrational and gamma-unstable or prolate deformed nuclei and is based on a method that was successfully applied before in the context of critical point symmetries.
\end{abstract}


\maketitle

\section{Introduction}

Critical point symmetries, as the E(5) and X(5) \cite{E5, X5} solutions of the Bohr Hamiltonian became known, have for more than a decade described the properties of transitional nuclei and were followed by a variety of similar models giving successful, parameter-free predictions of energy ratios and B(E2) ratios. In one of these models \cite{Dav}, a Davidson potential $(u(\beta)=\beta^2+\beta_0^4/\beta^2)$ is used, instead of the infinite square well in $\beta$, which is employed in both E(5) and X(5), and the critical point is determined by the procedure described in the next section. The results obtained resemble very closely those of E(5) and X(5).
 
The recent version of the Bohr Hamiltonian \cite{DDM}, where the mass term is allowed to depend on the $\beta$ variable of nuclear deformation is solved analytically using a Davidson potential in $\beta$ and by employing techniques from supersymmetric quantum mechanics \cite{SUSYQM}. In addition to the new set of analytic solutions, the newly introduced Deformation-Dependent Mass (DDM) model offers a remedy to the problematic behaviour of the moment of inertia in the Bohr Hamiltonian, where it appears to increase proportionally to $\beta^2$. In the DDM model the moments of inertia may increase at a lower rate, in agreement with experimental data. 

Recently, a solution of the DDM model with a Kratzer potential has also been obtained \cite{Kra}. The fact that the numerical results for the Davidson and the Kratzer potentials in the DDM framework are in general of the same quality, even though different functions for the dependence of the mass on the deformation are used in each case,  further supports the idea of a deformation dependent mass.

The purpose of this work is mainly to investigate the behaviour of the model parameters when the approach followed previously \cite{Var} for the study of shape transitions is applied.  

\section{An ``extremum'' approach to shape transitions }

The Davidson potential was among the first potentials proposed to describe shape changes in nuclei in a framework similar to that of the E(5) and X(5) solutions. Although, the Bohr Hamiltonian is solved exactly with a Davidson potential, its mere presence is not enough for a description of nuclei at the critical point, unless it is combined with the ``variational'' procedure, introduced in \cite{Dav, Var}. 

In this case, the energies are functions of the angular momentum L and of the parameter $\beta_0$ (the position of the potential minimum). Upon variation of $\beta_0$ (from 0 to sufficiently large values) the energy spectra change from those of the spherical type to those of the $\gamma$-unstable type. With the addition of a harmonic oscillator potential term around $\gamma=0$ and the approximation followed in X(5) the resulting spectra cover also the region from the spherical to the prolate-deformed nuclei.

Then, the critical point is identified as the value of $\beta_0$ that maximizes the rate of change of each energy ratio $(R_{L}=E(L)/E(2))$, in accordance with the observation that in a phase transition certain characteristic quantities change most abruptly. Therefore, one looks for the value of $\beta_0$ for which the first derivative of $R_{L}$ (for each separate value of L) with respect to $\beta_0$, becomes maximum and subsequently one uses these values to calculate the ``critical'' energy ratios. The method is reminiscent of the Variable Moment of Inertia (VMI) model, where an equilibrium condition ${\partial E }/{\partial J}|_{L=const}=0$ is also employed to determine the moment of inertia $J$  \cite{VMI}. 

As shown below, this method can be, almost trivially, extended in the DDM framework for the two cases of shape transitions, with the addition of an extra parameter. More specifically, the energy spectrum of the ground state band, in the DDM model with the Davidson potential \cite{DDM}, is given by the general expression:
\begin{equation}\label{eq:solve}
\epsilon_0= a \frac{29}{4}+\frac{1}{2}\sqrt{a^2+4k_1}+\frac{a}{2}\sqrt{1+4k_{-1}} + \frac{1}{4}\sqrt{(a^2+4k_1)(1+4k_{-1})}+a\Lambda
\end{equation}
where
\begin{eqnarray}
&k_1&=2+a^2[5(1-\delta - \lambda)+(1-2\delta)(1-2\lambda)+6+\Lambda]\nonumber \\
&k_0&=a[5(1-\delta - \lambda)+8+2\Lambda]\\
&k_{-1}&=2+\Lambda + 2\beta^4_0.
\end{eqnarray}

$\Lambda$ originates in the angular part of the original Schr\"odinger equation and is, thus a function of the angular momentum, taking a different form in the $\gamma$-unstable and deformed cases.  
It should be noted that for the purposes of our work, $\delta=\lambda=0$ in the above expressions.

\begin{figure}[!h]
\centering
\includegraphics[scale=0.9]{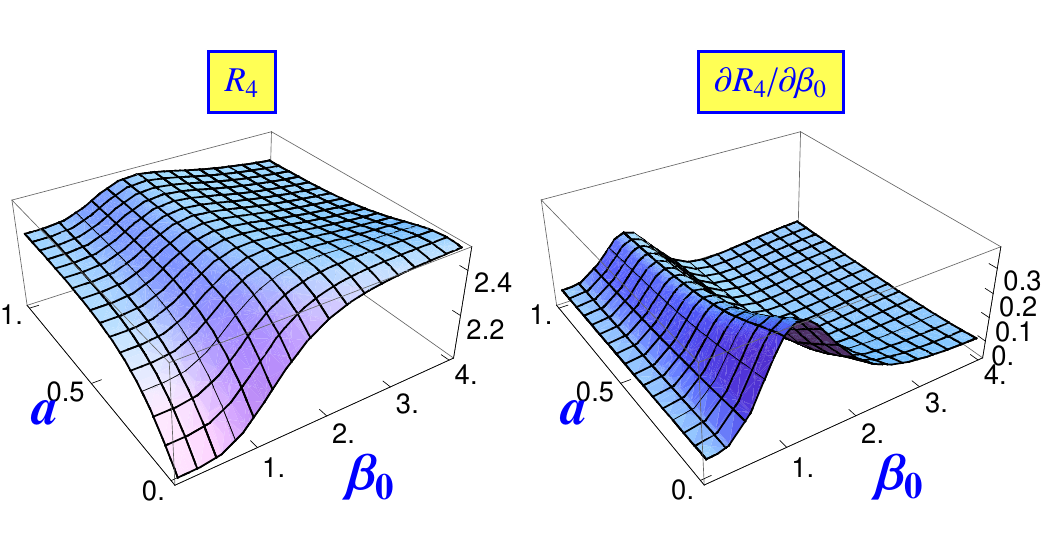}
\caption{$R_4$ energy ratio surface (left) and its derivative with respect to $\beta_0$ (right) as functions of $a$ and $\beta_0$ for the  $\gamma$-unstable case. The deformed case is very similar and is not shown.
}\label{fig1}
\end{figure}

As can be seen, the energies, apart from the angular momentum L (through $\Lambda$) and $\beta_0$, depend also on $a$, which is the extra  parameter that enters the formula of the mass as a function of deformation (\cite{DDM}). Consequently, the energy is represented graphically by a surface (fig. (\ref{fig1})), instead of a curve and the extremum condition mentioned above is implemented by finding the pairs of  $(a,\beta_0)_{crit}$ values that maximize the partial derivatives of $R_L$ with respect to $\beta_0$, for each value of L separately, if one wants to keep a close analogy with the original work presented in \cite{Dav}. A proper rescaling of the potential, like the one followed in \cite{Kra} may be necessary in order to lower the obtained $\beta_0$'s to more physical values. 

It should be noted that in the case of the DDM with a Davidson potential a physical interpretation of $a$ has been given, relating it to a curvature in the 5D space of the Bohr Hamiltonian and connecting it to the 6D space of the Interacting Boson Model (IBM) providing relevant interpretations of $a$ in each of the three IBM limits. A full discussion of these points can be found in \cite{Panos} and \cite{Aparam}.

In the $\gamma$-unstable case, $\Lambda=\tau(\tau+3)$ gives the eigenvalues of the second-order Casimir operator of SO(5), with $\tau$, the seniority quantum number, characterizing the irreducible representations of SO(5). The problem of what values of angular momentum L correspond to each value of $\tau$ is the group-theoretical problem of the SO(5)$\supset$SO(3) decomposition and is described in \cite{IBM, Wilets}. In the ground state band the correspondence takes the very simple form $L=2\tau$.

As shown in figure \ref{fig1} for the $R_4$ ratio, it increases monotonically with both $a$ and $\beta_0$ until it reaches asymptotically a certain value. For fixed values of $a$ the dependence on $\beta_0$ exhibits an inflection point, where the derivative (with respect to $\beta_0$) becomes maximum. This trend is very clear for values of $a$ close to zero and becomes smoother for increasing $a$. This can also be seen in the nearby plot of the partial derivative $\partial R_L/\partial \beta_0$ surface.
  
\begin{figure}[!h]
\centering
\includegraphics[scale=0.4]{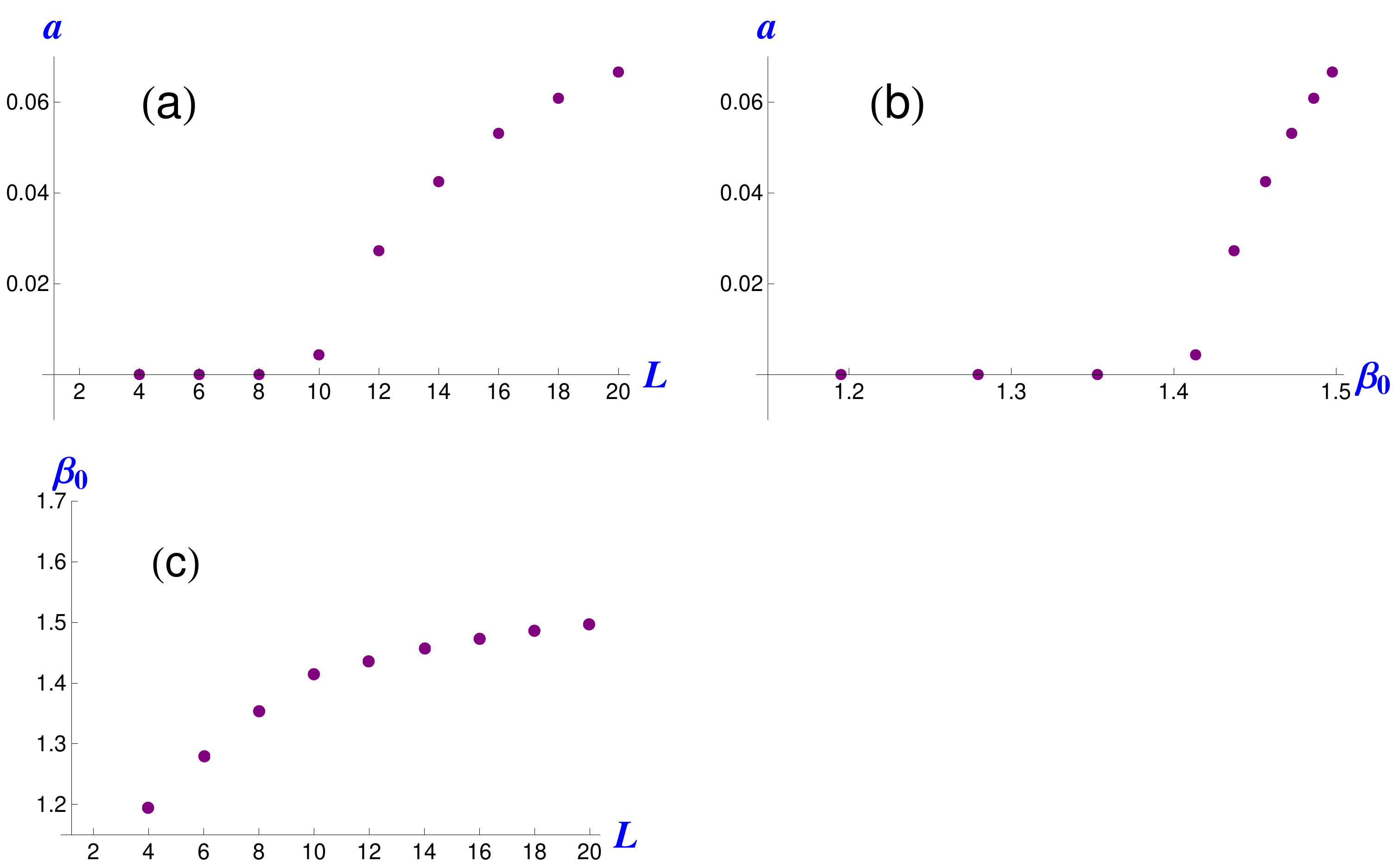}
\caption{Critical values for $a$ and $\beta_0$ for the various L values in the `spherical to  $\gamma$-unstable' transition.
}\label{fig2}
\end{figure}

An interesting aspect is revealed when one considers the evolution of the critical values of the parameters $a$ and $\beta_0$ as L increases.  As can be seen from fig.(\ref{fig2}a), $(a)_{crit}$ values remain zero for small angular momenta and jump to some finite value at L=10. Such a behaviour is reminiscent of a phase transition \cite{Meso}, with L playing the role of control parameter. This abrupt change however is not due to some discontinuity in the $\partial R_L/\partial \beta_0$ surface. As $a$ is, by definition, $a\geq 0$ \cite{DDM}, $(a)_{crit}=0$ represents the constrained maximum of the $\partial R_L/\partial \beta_0$ surface for low values of L. At the same time, $(\beta_0)_{crit}$ shows a linear increase with L up to $L=10$, after which it continues to grow linearly at a lower rate. It should be noted that for $a=0$ the results are identical to those obtained in ref. \cite{Var}.

\begin{figure}[!h]
\centering
\includegraphics[scale=0.4]{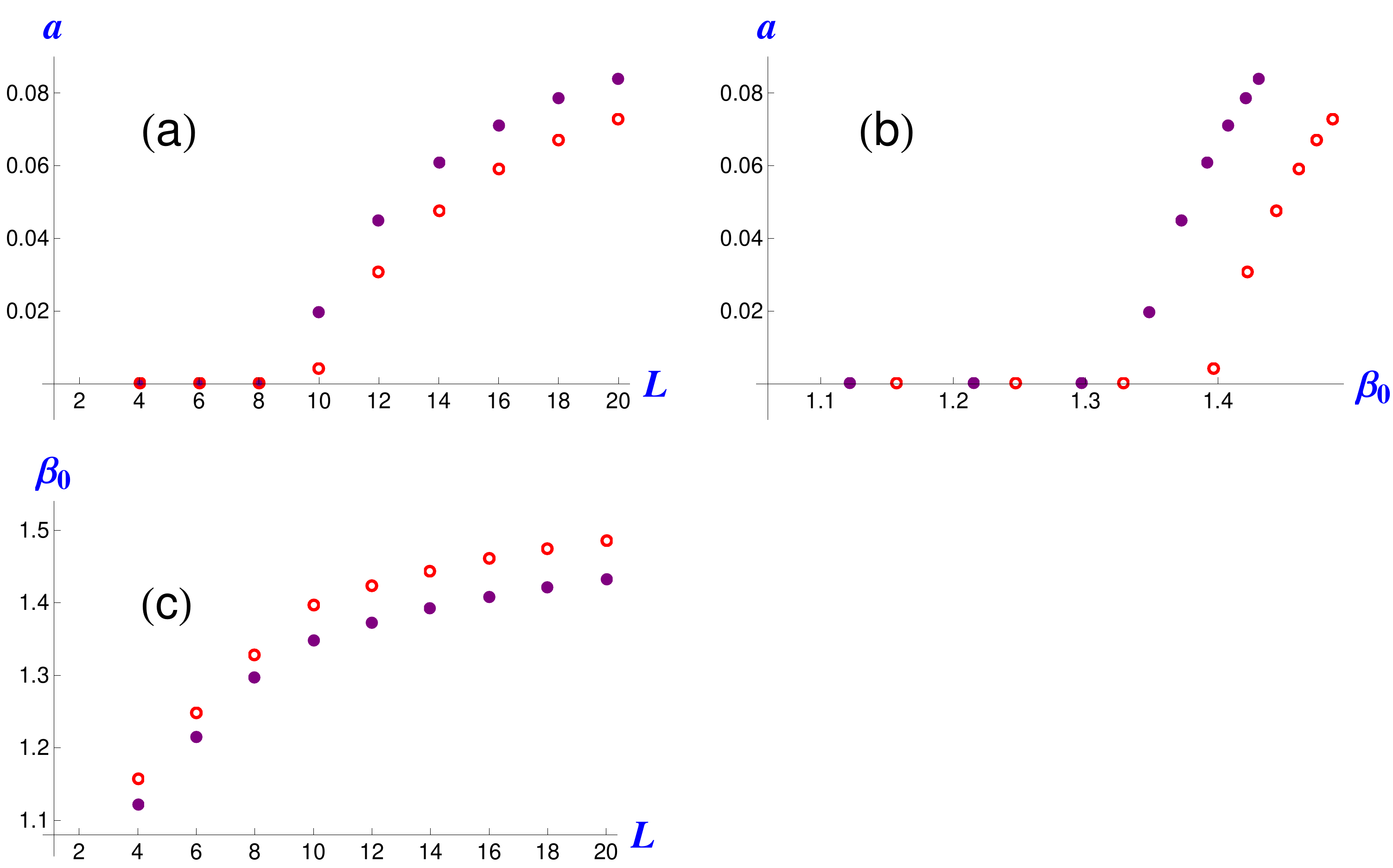}
\caption{Critical values for $a$ and $\beta_0$ for the various L values in the `spherical to deformed' transition. Dots correspond to $c=0$ and open circles to $c=0.1$.
}\label{fig3}
\end{figure}

A potential that can describe axially (prolate) deformed nuclei and still allows exact separation of variables in the Bohr Hamiltonian is of the form \cite{Wilets,ESD,Fort1,Fort2} 
\begin{equation}
v(\beta,\gamma)=u(\beta)+\frac{w(\gamma)}{\beta^2}
\end{equation}
where $ u(\beta)$ involves the Davidson potential and 
\begin{equation}
w(\gamma)=\frac{1}{2}(3c)^2\gamma^2
\end{equation}
represents a  harmonic oscillator centered around $\gamma=0$. 
The use of this in the DDM framework yields for $\Lambda$ the expression 
\begin{equation}
\Lambda=\epsilon_{\gamma}-\frac{K^2}{3}+\frac{L(L+1)}{3},
\end{equation}
where  $\epsilon_{\gamma}=6c(n_{\gamma}+1)$ and $n_{\gamma}=0,1,2,...$ is the number of $\gamma$-oscillation quanta. For the ground state band which we examine here, we have $K=0$ and $n_{\gamma}=0$. The cases $c=0$ and $c=0.1$ are examined (fig. \ref{fig3}) and the same general trend of the parameters is observed as in the $\gamma$-unstable case.

\begin{figure}[!h]
\centering
\includegraphics[scale=0.32]{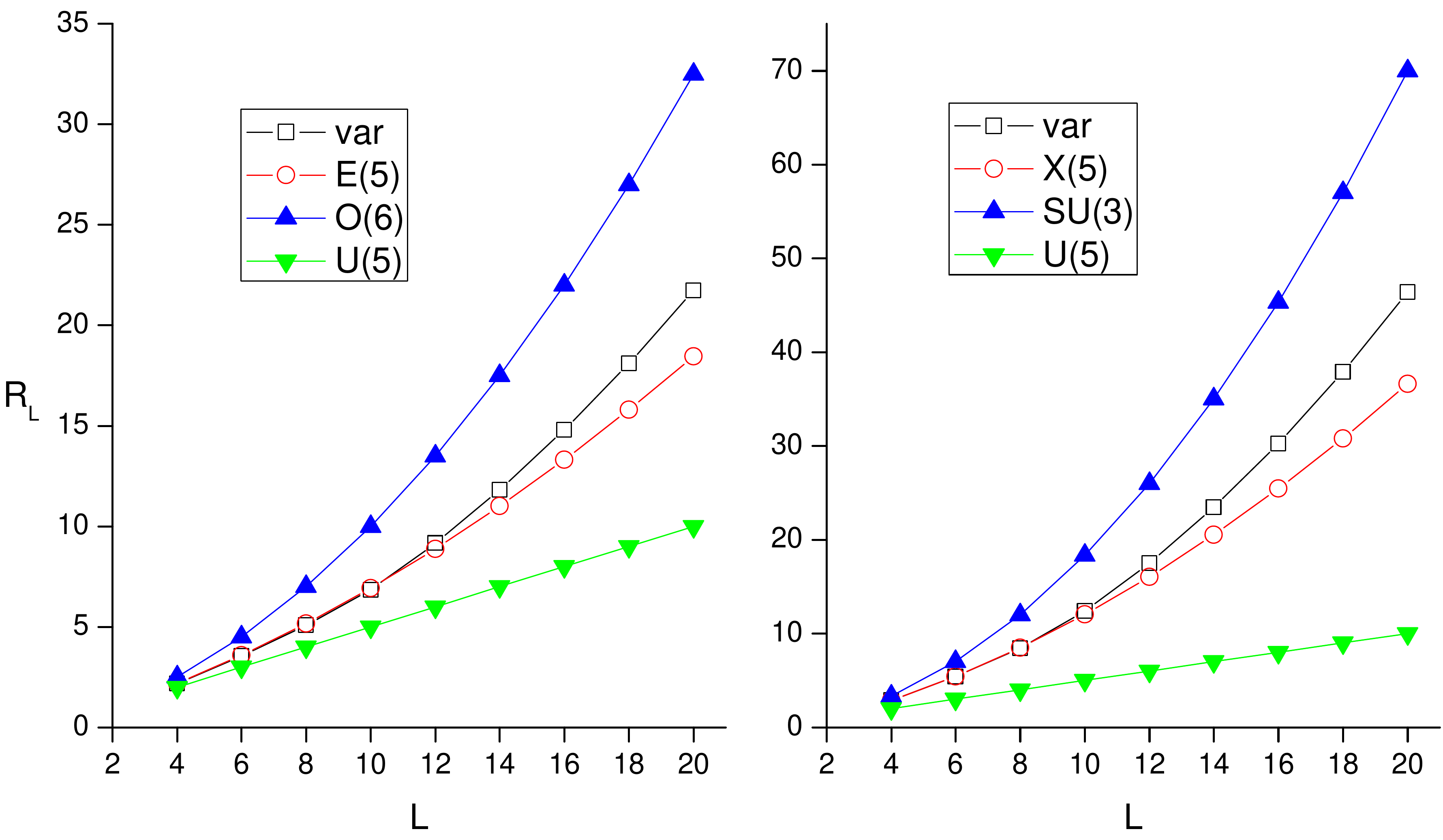}
\caption{Critical $R_L$ energy ratios (var) as functions of L for the ground state band for the spherical to $\gamma$-unstable (left) and the spherical to deformed (right) transitions, compared to the E(5) and X(5) results respectively.
}\label{fig4}
\end{figure}

The results of this procedure where $a$ is variable are shown in fig. \ref{fig4}. The same method can be applied for fixed values of the $a$ parameter. Specifically, it has been found that the extremum condition for $a=0.011$ and $a=0.0035$ reproduces the ground band $R_L$ ratios for E(5) and X(5), respectively with very good accuracy, as table 1 demonstrates.  
\newpage 
\begin{table*}

\caption{Critical $R_L$ energy ratios for the ground state band for fixed values of $a$, compared to the E(5) and X(5) results.  }

\begin{tabular}{ l | c c | c c}
\hline 

L & $a=0.011$ & E(5) & $a=0.0035$ & X(5) \\

\hline

4 & 2.19663 & 2.199 & 2.90566 & 2.904 \\
6 & 3.58597 & 3.59 & 5.43696 & 5.43 \\
8 & 5.16385 & 5.169 & 8.49561 & 8.483 \\
10 & 6.92715 & 6.934 & 12.0404 & 12.027 \\
12 & 8.87378 & 8.881 & 16.051 & 16.041 \\
14 & 11.0025 & 11.009 & 20.5167 & 20.514 \\
16 & 13.3126 & 13.316 & 25.4323 & 25.437 \\
18 & 15.8041 & 15.799 & 30.7955 & 30.804 \\
20 & 18.477 & 18.459 & 36.6061 & 36.611 \\

\hline
\end{tabular}
\end{table*}


\section{Moments of inertia in the Deformation Dependent Mass model.} 
\begin{figure}[!h]
\centering
\includegraphics[scale=0.45]{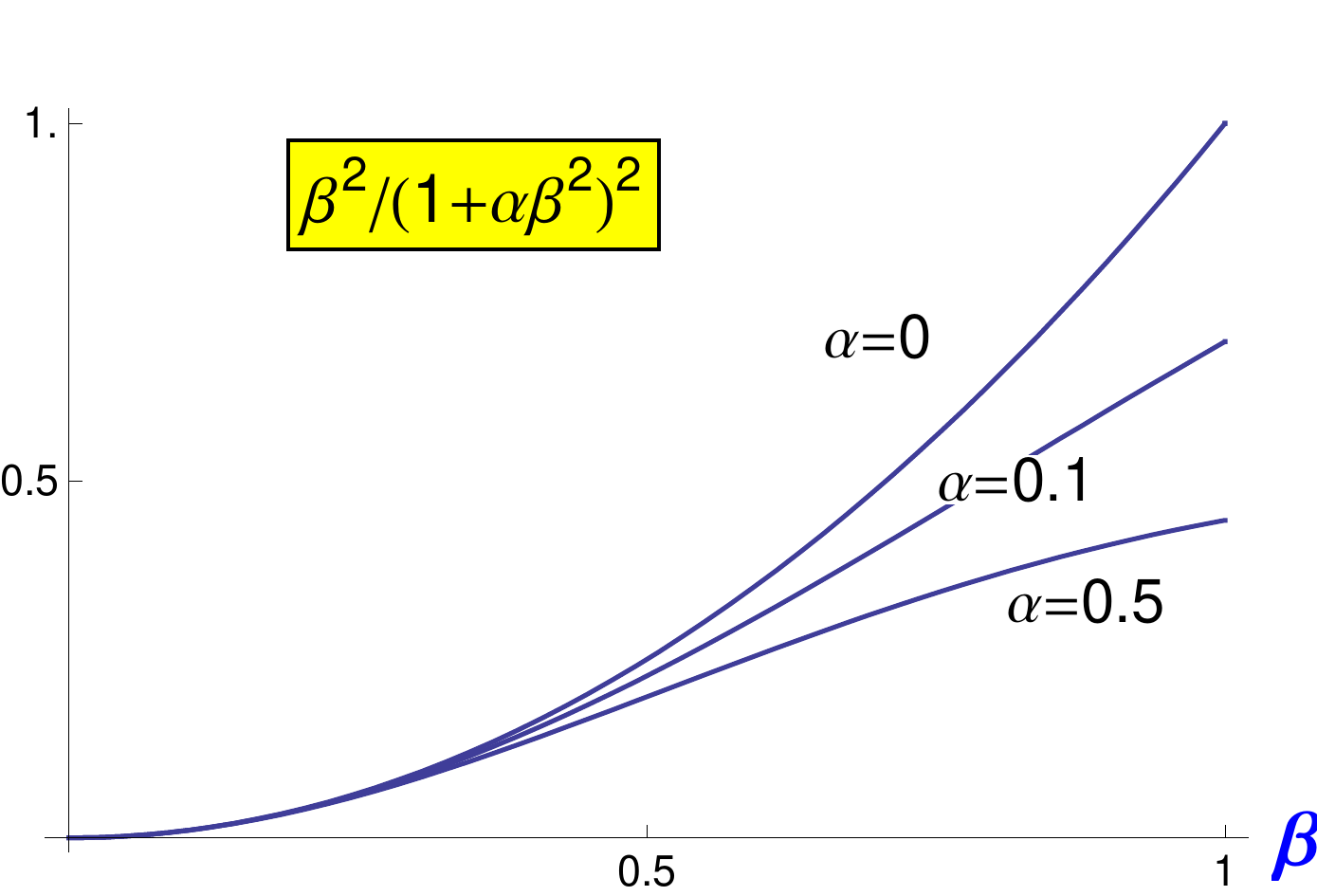}
\caption{$\beta$-dependence of the moment of inertia for the case of the Davidson potential for various values of the $a$ parameter.
}\label{fig5}
\end{figure}

As already mentioned, the Deformation Dependent Mass model allows for a moderation in the rate of increase of the moment of inertia from the undesirable $\beta^2$ dependence. This can be seen in fig. \ref{fig5}, where the moment of inertia for the case of the Davidson potential has been plotted as a function of $\beta$.  

\begin{figure}[!h]
\centering
\includegraphics[scale=0.4]{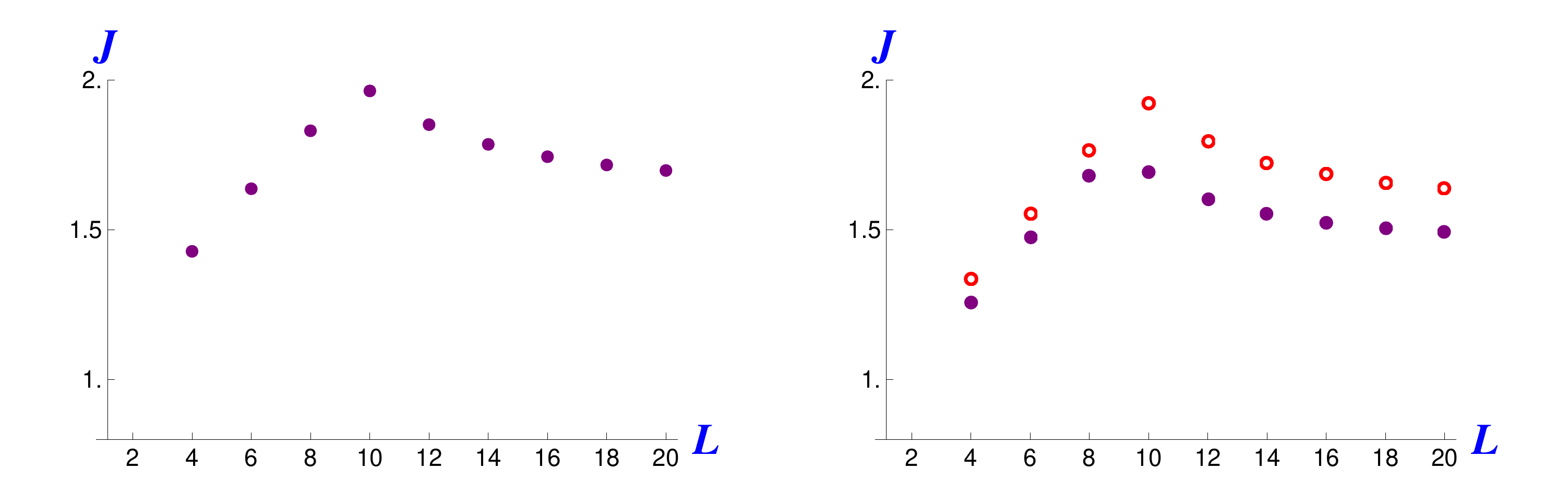}
\caption{Moments of inertia for the critical values of $a$ and $\beta_0$ obtained in the `spherical to $\gamma$-unstable' transition (left) and the `spherical to deformed' transition (right). In the right panel, dots correspond to $c=0$ and open circles to $c=0.1$.
}\label{fig6}
\end{figure}

An interesting behaviour is observed if the calculated critical values of the $a$ and $\beta_0$ parameters are used in the formula for the moment of inertia:
\begin{equation}
J=\frac{\beta^2}{(1+a \beta^2)^2}
\end{equation}
As can be seen in figure \ref{fig6}, the changes in the critical values with angular momentum are reflected in the respective values of the moments of inertia, making them  increase linearly up to L=10 and decrease thereafter, when $a$ starts taking non-zero values. This ``downbending'' behaviour, although not usual, has been observed and is mentioned in \cite{VMI2}. It should be noted, however, that downbending in the present treatment is derived from the spectrum of only one band, the ground band. \\

\section{Conclusion}
The Deformation-Dependent Mass (DDM) model solves not only the problem of unphysical moments of inertia in the original Bohr Hamiltonian, but it provides it with a richer structure, with the introduction of an extra parameter $a$. For fixed values of $a$ and the use of a previous method to obtain the critical points within the Bohr Hamiltonian, one can reproduce the E(5) and X(5) results, while interesting trends are observed if one leaves $a$ to vary. These trends are the subject of further study. 

\section{Acknowledgements}

Support by the Bulgarian National Science Fund (BNSF) under Contract No. DFNI-E02/6 is gratefully acknowledged. D. Petrellis acknowledges additional financial support by the Scientific Research Projects Coordination Unit of Istanbul University under Project No 50822.


\section*{References} 

\begin{thebibliography}{99}

\bibitem{E5}
F. Iachello, Phys. Rev. Lett. {\bf 85}, 3580 (2000).

\bibitem{X5}
F. Iachello, Phys. Rev. Lett. {\bf 87}, 052502 (2001).

\bibitem{Dav}
D. Bonatsos, D. Lenis, N. Minkov, D. Petrellis, P.P. Raychev and P.A. Terziev, Phys. Lett. B {\bf 584}, 40 (2004).

\bibitem{DDM}
D. Bonatsos, P. E. Georgoudis, D. Lenis, N. Minkov, and C. Quesne, Phys. Rev. C {\bf 83}, 044321 (2011).

\bibitem{SUSYQM}
F. Cooper, A. Khare, and U. Sukhatme, Supersymmetry in Quantum Mechanics (World Scientific, Singapore, 2001).

\bibitem{Kra}
D. Bonatsos, P. E. Georgoudis, N. Minkov, D. Petrellis and C. Quesne, Phys. Rev. C {\bf 88}, 034316 (2013).

\bibitem{Var}
D. Bonatsos, D. Lenis, N. Minkov, D. Petrellis, P.P. Raychev and P.A. Terziev, Phys. Rev. C {\bf 70}, 024305 (2004).

\bibitem{VMI}
M. A. J. Mariscotti, G. Scharff-Goldhaber and B. Buck, Phys. Rev. {\bf 178}, 1864 (1968).

\bibitem{Panos}
P.E. Georgoudis, Phys. Lett. B {\bf 731}, 122 (2014).

\bibitem{Aparam}
D. Bonatsos, N. Minkov and D. Petrellis, J. Phys. G: Nucl. Part. Phys. {\bf 42}, 095104 (2015).

\bibitem{IBM}
F.\ Iachello and A. Arima, \textit{The Interacting Boson Model} (Cambridge University Press, Cambridge, 1987).

\bibitem{Wilets}
L. Wilets and M. Jean, Phys. Rev. {\bf 102}, 788 (1956).

\bibitem{Meso}
F. Iachello and N. V. Zamfir, Phys. Rev. Lett. {\bf 92}, 212501 (2004).

\bibitem{Fort1}
L. Fortunato, Phys. Rev. C {\bf 70}, 011302 (2004).

\bibitem{Fort2}
L. Fortunato, Eur. Phys. J. A {\bf 26}, 1 (2005).

\bibitem{ESD}
D. Bonatsos, E. A. McCutchan, N. Minkov, R. F. Casten, P. Yotov, D. Lenis, D. Petrellis, and I. Yigitoglu, Phys. Rev. C {\bf 76}, 064312 (2007).

\bibitem{VMI2}
G.Scharff-Goldhaber, C.B.Dover and A.L.Goodman, Ann .Rev. Nucl. Sci. {\bf 26}, 239 (1976).

\end{thebibliography}
\end{document}